\documentclass{article}

\usepackage[final]{neurips_2022_ml4ps}

\usepackage[utf8]{inputenc} 
\usepackage[T1]{fontenc}    
\usepackage{hyperref}       
\usepackage{url}            
\usepackage{booktabs}       
\usepackage{amsfonts}       
\usepackage{nicefrac}       
\usepackage{microtype}      
\usepackage{xcolor}         
\usepackage{natbib}
\usepackage{graphicx}

\title{Identifying AGN host galaxies with \\convolutional neural networks}

\author{%
  Ziting Guo\\
  Yale-NUS College\\
  \#01-220, 16 College Avenue West \\  
  Singapore 138527 \\
  \texttt{guo.ziting@aya.yale.edu} \\
  
   \And
   John F. Wu \thanks{Secondary affiliation: Department of Physics \& Astronomy, Johns Hopkins University, 3400 N. Charles St., Baltimore, MD 21218} \\
   Space Telescope Science Institute \\
   3700 San Martin Drive \\
   Baltimore, MD 21218 \\
   \texttt{jowu@stsci.edu} \\
   \AND
   Chelsea E. Sharon \\
   Yale-NUS College\\
   \#01-220, 16 College Avenue West \\  
   Singapore 138527 \\
   \texttt{chelsea.sharon@yale-nus.edu.sg} \\

}

\begin{document}

\maketitle

\begin{abstract}
Active galactic nuclei (AGN) are supermassive black holes with luminous accretion disks found in some galaxies, and are thought to play an important role in galaxy evolution. However, traditional optical spectroscopy for identifying AGN requires time-intensive observations. We train a convolutional neural network (CNN) to distinguish AGN host galaxies from non-active galaxies using a sample of 210,000 Sloan Digital Sky Survey galaxies. We evaluate the CNN on 33,000 galaxies that are spectrally classified as composites, and find correlations between galaxy appearances and their CNN classifications, which hint at evolutionary processes that effect both galaxy morphology and AGN activity. With the advent of the Vera C. Rubin Observatory, \textit{Nancy Grace Roman Space Telescope}, and other wide-field imaging telescopes, deep learning methods will be instrumental for quickly and reliably shortlisting AGN samples for future analyses.
\end{abstract}

\section{Introduction}

Nearly all galaxies are theorized to contain a supermassive black hole at the galactic center, but only in certain galaxies do interstellar gas and dust accumulate around the supermassive black hole to form a bright, disk-shaped structure called active galactic nucleus (AGN). Most galaxy evolution models suggest that AGN regulates the rate of star formation in a galaxy.

As AGN are structurally small relative to their host galaxies, the identification of AGN traditionally relies on expensive spectroscopic analysis. The BPT diagram \citep{BPT} plots the [O\textsc{iii}]/${\rm H}\beta$ emission line flux ratio against that of [N\textsc{ii}]/${\rm H}\alpha$, to distinguish AGN from non-active galaxies. In between active and non-active galaxies are the spectral composites with intermediate emission-line ratios, such as ordinary AGN whose spectra may have been contaminated by nearby star-forming regions \citep{Ho1993}. While essential for AGN identification, spectroscopy takes $\sim 10^3$ times more time than imaging. It may be possible to identify AGN based on galaxy images alone, given AGN's connection to star formation. Galaxy evolution is theorized to depend strongly on the co-evolution between the central supermassive black hole and the rest of the galaxy, and AGN presence may therefore be linked to galaxy morphology. \cite{Kauffmann}'s study on galaxies from the Sloan Digital Sky Survey (SDSS) reveals patterns in the physical properties of AGN host galaxies, showing promise for an image-based AGN identification method. 

Astronomers have already begun employing data-driven methods to search for AGN in wide-field imaging surveys. For example, \citet{tabular_agn} identify various classes of AGN candidates using catalogues of optical and near-infrared photometry.
Another work found that optically obscured (Type I) and unobscured (Type II) AGN can be distinguished from each other using features derived from X-ray observations \citep{xray_agn}.
Some AGN are variable, and \citet{quasar_cnn} train a convolutional neural network to identify quasar candidates from photometric time-series data.
Finally, \citet{predicting_ugc2885} have trained a CNN to predict galaxy spectra directly from image cutouts \citep{spectra_dl}, and have estimated the BPT classification of a galaxy by extracting lines flux ratios from the CNN-derived spectrum.
We present results that are most similar to this last work, except that we directly estimate BPT classifications from image cutouts.

In this work, we show that a CNN can detect the presence of AGN from its host galaxy's appearance. 
We evaluate our trained CNN on galaxies spectroscopically labelled as composite, in order to determine whether they are more likely to be star-forming non-active objects, or AGN hosts.
The results support our physical intuition: smaller, bluer composite systems are more often predicted to be star-forming, while redder bulge-dominated composite systems are more often predicted to host AGN. Using only broadband imaging, our method can be used to efficiently prioritize AGN candidates for follow-up spectroscopy in future imaging-only sky surveys, which will capture more galaxy images than previous sky surveys combined.

\section{Methodology} \label{method}

We select galaxy images from the SDSS Data Release 8 (DR8) spectroscopic catalog \citep{SDSS_MPA_JHU_1, SDSS_MPA_JHU_2, Kauffmann}. We download $gri$-band $160 \times 160$-pixel SDSS image cutouts at a pixel scale of $0.262^{\prime\prime}$ from the Legacy Survey website \citep{desi}.  We require that galaxies in our sample be similar in size in final images by restricting galaxies to spectroscopic redshift 0.02 $< z < 0.3$. Following the selection criteria from \cite{Kauffmann}, each of the four emission lines, H$\alpha$, [N\textsc{ii}], H$\beta$, [O\textsc{iii}], must have a non-zero flux with signal-to-noise ratio $(S/N) > 3$. Our ground truth labels of whether a galaxy contains AGN were obtained by BPT-based spectroscopic analysis in \cite{Kauffmann}; we note that these classifications are model-dependent and only represent an estimate of the ground truth. 

Composite and unclassifiable galaxies are excluded from training and validation. Of the remaining 210,332 galaxies, $\sim21\%$ have AGN. Following a 80-20 train-validation split, we augment images in the training set to improve our model's robustness. A good CNN should be able to make accurate predictions despite scatter in the central galaxy's location, size and viewing angle. We stochastically apply an augmentation pipeline consisting of rotation, reflection, warping, and resizing to galaxy images in the training set, for which the augmentation degrees were optimized in preliminary experiments.

Our baseline model is \texttt{resnet18}, an 18-layer residual CNN, with ReLU replaced by Mish \citep{mish}. With a batch size of 64, and \texttt{ranger} \citep{ranger}, a combination between \texttt{rectified Adam} \citep{radam} and \texttt{lookahead} \citep{lookahead}, as optimizer, the model learns the relationship between galaxy morphology and AGN presence by minimizing the cross-entropy loss between its predictions and ground truth labels. Learning rates are scheduled by \texttt{fastai}'s default learning rate scheduler, \texttt{1cycle policy} \citep{1cycle}. These hyperparameter choices have shown to speed up convergence and improve accuracy in other astrophysical computer vision tasks \citep{xSAGA}.

\section{Results} \label{res}

\subsection{CNN classifications on active and non-active galaxies}

We report that \texttt{resnet18} classifies active and non-active galaxies with $0.894 \pm{0.001} $ accuracy, $0.738 \pm{0.020}$ precision, $0.765 \pm{0.046} $ recall, and $0.750 \pm{0.012}$ F1-score. Uncertainties are determined by repeating the methodology using five different random seeds. Our experiments confirm the usefulness of morphological information in identifying active galaxies, and in turn, the connection between galaxy morphology and AGN presence.

\begin{figure}[t]
    \centering
    \includegraphics[width=\textwidth]{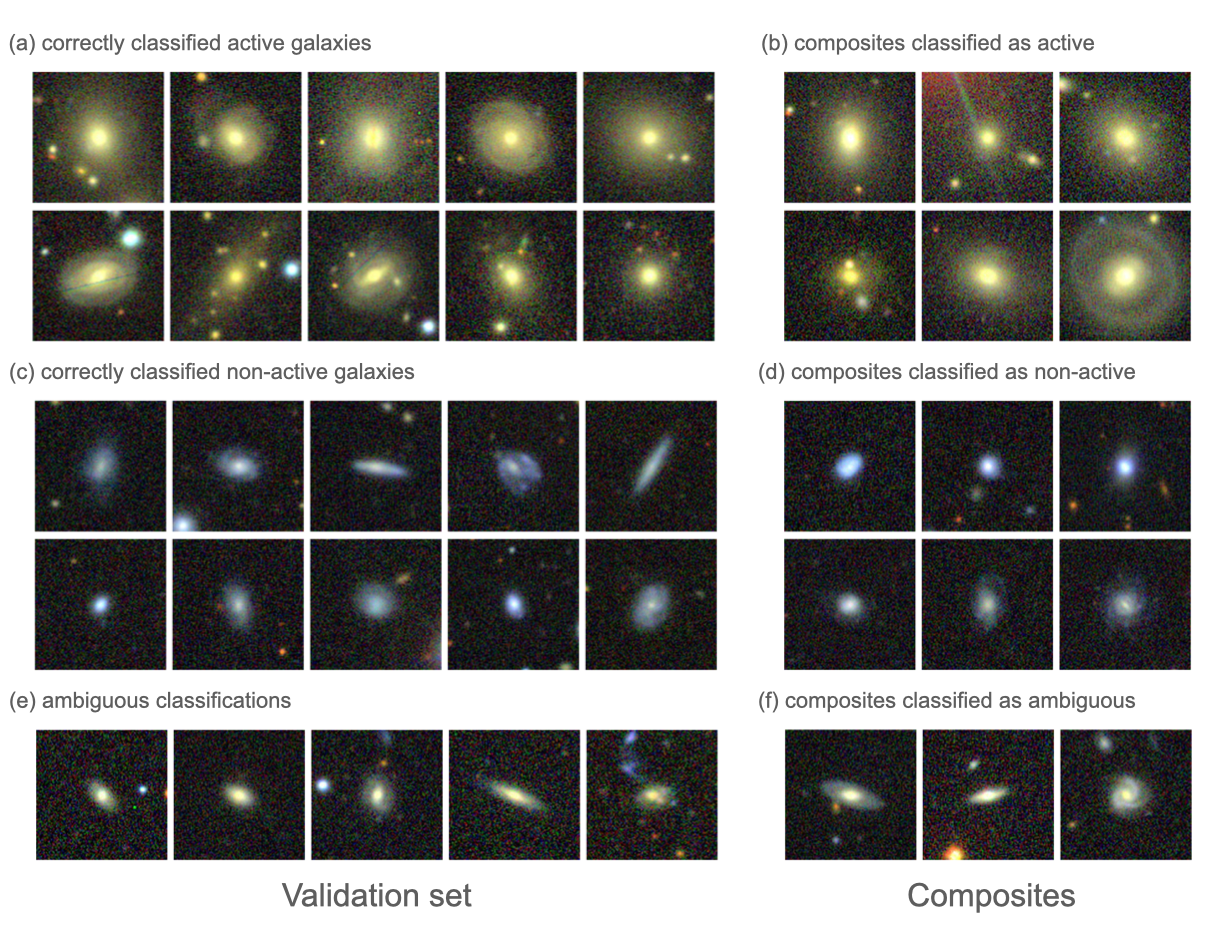}
    \caption{
    Spectroscopically classified AGN hosts and non-active galaxies from the validation set (\textit{left}), and composites (\textit{right}), as defined in \cite{Kauffmann}.
    Panel (a), \textit{left} shows ground truth active galaxies which have been classified as active by the CNN, and Panel (b), \textit{left} shows correctly classified non-active galaxies. Panels (a) and (b), \textit{right} are composite galaxies which have been classified as active and non-active respectively. 
    Panel (c), \textit{left and right} shows ambiguous classifications, for which CNN predicts equal probability for that it is active and that it is non-active. The small, reddish ambiguous objects in Panel (c) have features intermediate between active and non-active classifications.
    }
    \label{fig:image}
\end{figure}

Examples of typical active and non-active galaxies as identified by CNN are displayed in Figure \ref{fig:image}, left. Panels (a) and (b), left show correctly classified active and non-active galaxies respectively.
There is a clear distinction in morphology between what the model perceives as a typical AGN host and what as a typical non-active galaxy. Correctly classified AGN host galaxies in Figure \ref{fig:image}(a) are redder and bigger (both in apparent and physical sizes) with dominant central bulges, which is consistent with \cite{Kauffmann}'s observation that nearly all dust-unobscured AGN reside in massive galaxies with little star formation. On the other hand, galaxies correctly classified as non-active are smaller star-forming galaxies with blue bulges. 
Among objects with highest-confidence CNN predictions, i.e. correct classifications with 10\% least cross entropy losses, correctly classified active galaxies have median color index $g - i = 1.450$ and Petrosian $r$-band half-light radius $R_{50} = 3.195$ arcsec, whereas correctly classified non-active galaxies have median $g - i = 0.698$ and $R_{50} = 2.778$ arcsec.
These visual patterns are consistent with the black hole mass - luminosity relation \citep{bh_mass_lumin} and the mass-metallicity relation \citep{mass_metal}.

Despite the presence of unrelated objects and artefacts in galaxy images in Figure \ref{fig:image}, the CNN learns to make its predictions for the central galaxies, invariant to other unrelated objects in the field of view.

\subsection{Using the trained CNN to classify composite galaxies}

One of the key advantages of a machine learning based AGN identification method is to classify composite galaxies that traditional spectroscopic analyses fail to differentiate. Examples of composite galaxies likely to be active and non-active respectively, as well as those ambiguous to the CNN, are displayed in Figure \ref{fig:image}, right.

We evaluate our trained CNN on the dataset of spectroscopic composites, while keeping in mind that composite galaxies are not represented at all in our training set. Unsurprisingly, composite galaxies classified as AGN by the CNN in Figure \ref{fig:image} (b) greatly resemble the typical AGN in Figure \ref{fig:image}(a). Similarly, both typical non-active galaxies, and the composite galaxies identified to be non-active by the CNN (Figure \ref{fig:image} (c) and (d) respectively), lack dominant central bulges. Compared to predicting active and non-active galaxies, our CNN is less confident in predicting composite galaxies, suggesting that spectroscopically determined composites are also morphological intermediates.

\section{Discussion}

\begin{figure}[t]
    \centering
    \includegraphics[width=0.7\textwidth]{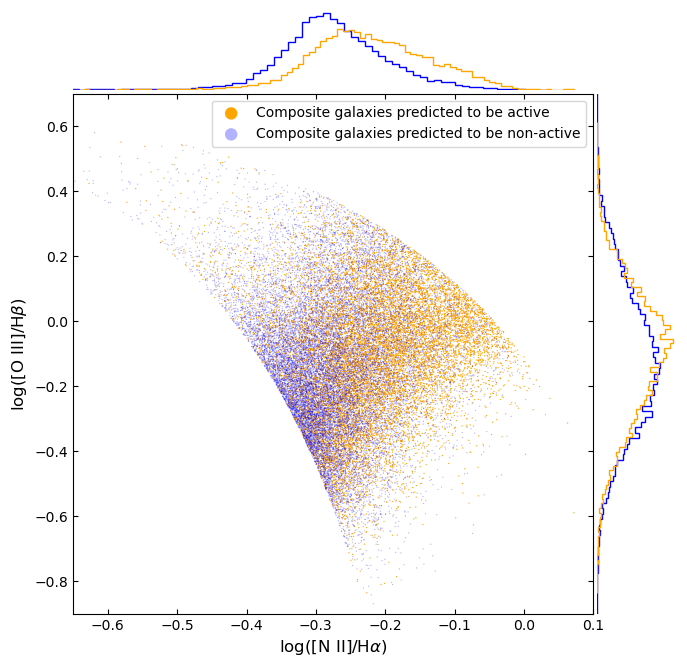}
    \caption{
    CNN predictions on galaxies spectroscopically classified as composites are shown on the BPT diagram in the main panel. Normalized histograms of composite galaxies' [O\textsc{iii}]/${\rm H}\beta$ and [N\textsc{ii}]/${\rm H}\alpha$ line ratios are shown in the right and top panels respectively. Orange dots and lines represent spectroscopic composites the CNN identifies as active, whereas blue ones are classified as non-active. The central locus is bound by the two parabolas in \cite{kewley_starburst} and \cite{Kauffmann}.
    }
    \label{fig:bpt_composite}
\end{figure}

Our work is the first application of CNN to identify AGN directly from optical images. We benchmark our CNN's performance against a baseline \texttt{xgboost} model trained on $gri$ photometric data that is similar to the approach by \cite{tabular_agn}. Again, we obtain the relevant photometric information from SDSS. Our \texttt{resnet18} trained on optical images outperforms the \texttt{xgboost} model in all chosen metrics (the best \texttt{xgboost} achieved 0.881 $\pm$ 0.001 accuracy, 0.726 $\pm$ 0.005 precision, 0.694 $\pm$ 0.007 recall, 0.710 $\pm$ 0.003 F1-score). \texttt{xgboost} performs on par with the CNN only when we include full $ugriz$ photometry and Petrosian radius. This comparison indicates that galaxy size, in addition to color, is also important for identifying AGN in our SDSS sample.

In Figure~\ref{fig:bpt_composite}, we show CNN classifications of spectroscopic composite galaxies in a BPT diagram.
A $2d$ Kolmogorov-Smirnov test reveals that CNN-predicted active and non-active galaxies have different distributions ($p < 10^{-3}$), confirming our expectation that objects classified as active lie closer to the right-hand side of the diagram (the region of spectroscopically defined AGN), and objects classified as non-active lie closer to the left-hand side.
We note that [N\textsc{ii}]/H$\alpha$ is also a tracer of gas-phase metallicity, and that the spectroscopic ``ground truth'' classifications are also metallicity-dependent.
It has also been shown that metallicity can be predicted directly from galaxy image cutouts \citep[][]{z_dl}, which supports our thesis that galaxies' morphologies are critical for identifying their physical properties.

Our deep learning approach can efficiently select AGN host galaxies from future imaging-only sky surveys. The Vera C. Rubin Observatory will go deeper and have higher resolution than SDSS. It will also detect many time-varying AGN at optical wavelengths. The \textit{Nancy Grace Roman Space Telescope} will have exquisite resolution, and survey many AGN host galaxies out to higher redshifts. It will have tremendous low-surface brightness sensitivity, enabling better morphological characterization of extended galaxy disks, tidal debris, and other interstellar gas and dust. Using a comprehensive, CNN-selected sample of AGN candidates, we astronomers will be able to robustly study the co-evolution of galaxies and their supermassive black holes. 

We have demonstrated that a CNN can identify AGN in low-redshift galaxies solely from optical imaging. We note that AGN hosts and non-active galaxies can overlap in the space of line flux ratios, particularly in the composite regime, but our machine learning technique hints at a way to separate probable AGN hosts from non-active galaxies on the basis of galaxy morphology. Future wide-field imaging surveys, such as the Legacy Survey of Space and Time by the Vera C. Rubin Observatory, and the High Latitude Wide Area Imaging Survey by the \textit{Nancy Grace Roman Space Telescope}, can benefit from efficient selection of AGN candidates (or other interesting phenomena) by using CNNs.

\section*{Broader Impact}

Like most deep learning projects, the process of training our model, required substantial computational resources and contributed to global warming. 

For the positive impact, by keeping our code open-source, we have continued to build on open science with archival data sets.

\section*{Acknowledgements and Disclosure} \label{ack}
Funding for SDSS-III has been provided by the Alfred P. Sloan Foundation, the Participating Institutions, the National Science Foundation, and the U.S. Department of Energy Office of Science. The SDSS-III web site is \url{http://www.sdss3.org/}.

SDSS-III is managed by the Astrophysical Research Consortium for the Participating Institutions of the SDSS-III Collaboration including the University of Arizona, the Brazilian Participation Group, Brookhaven National Laboratory, Carnegie Mellon University, University of Florida, the French Participation Group, the German Participation Group, Harvard University, the Instituto de Astrofisica de Canarias, the Michigan State/Notre Dame/JINA Participation Group, Johns Hopkins University, Lawrence Berkeley National Laboratory, Max Planck Institute for Astrophysics, Max Planck Institute for Extraterrestrial Physics, New Mexico State University, New York University, Ohio State University, Pennsylvania State University, University of Portsmouth, Princeton University, the Spanish Participation Group, University of Tokyo, University of Utah, Vanderbilt University, University of Virginia, University of Washington, and Yale University.

Code for this project can be found at \url{https://github.com/cherryquinnlg/agn-convnets}.

\raggedbottom
\bibliography{neurips_2022_bib}{}

\bibliographystyle{aasjournal}
\end{document}